\documentclass[pre, aps, showpacs, showkeys]{revtex4}
\usepackage{graphicx}
\begin{document}

\title{Modulated envelope localized wavepackets associated with  \\
electrostatic plasma waves \footnote{Proceedings of the
\textit{International Conference on Plasma Physics - ICPP 2004},
Nice (France), 25 - 29 Oct. 2004; contribution P3-044; available
online at: \texttt{http://hal.ccsd.cnrs.fr/ccsd-00001889/en/}  .}}

\author{Ioannis Kourakis\footnote{On leave from: U.L.B. -
Universit\'e Libre de Bruxelles, Physique Statistique et Plasmas
C. P. 231, Boulevard du Triomphe, B-1050 Brussels, Belgium; also:
Facult\'e des Sciences Apliqu\'ees - C.P. 165/81 Physique
G\'en\'erale, Avenue F. D. Roosevelt 49, B-1050 Brussels, Belgium;
\\Electronic address: \texttt{ioannis@tp4.rub.de}}
and Padma Kant Shukla\footnote{Electronic address:
\texttt{ps@tp4.rub.de}}} \affiliation{Institut f\"ur Theoretische
Physik IV, Fakult\"at f\"ur Physik und Astronomie,
Ruhr--Universit\"at Bochum, D-44780 Bochum, Germany}
\date{\today}

\begin{abstract}
The nonlinear amplitude modulation of known electrostatic plasma
modes is examined in a generic manner, by applying a collisionless
fluid model. Both cold (zero-temperature) and warm fluid
descriptions are discussed and the results are compared. The
moderately nonlinear oscillation regime is investigated by
applying a multiple scale technique. The calculation leads to a
Nonlinear Schrödinger-type Equation (NLSE), which describes the
evolution of the slowly varying wave amplitude in time and space.
The NLSE admits localized envelope (solitary wave) solutions of
bright- (pulses) or dark- (holes, voids) type, whose
characteristics (maximum amplitude, width) depend on intrinsic
plasma parameters. Effects like amplitude perturbation
obliqueness, finite temperature and defect (dust) concetration are
explicitly considered. The relevance with similar highly localized
modulated wave structures observed during recent satellite
missions is discussed.
\end{abstract}


\maketitle

\section{Introduction}

The \emph{amplitude modulation} (AM) of waves is a generic
nonlinear mechanism, which is long known to dominate finite
amplitude wave propagation in dispersive media. In a generic
context, the occurrence of AM is manifested as a slow variation of
the wave's amplitude in space and time, which may be due to
parametric wave coupling, interaction between high- and low-
frequency modes or self-interaction of the carrier wave
(\emph{auto-} or \emph{self-}modulation). The relation of this
phenomenon to effects like secondary \emph{harmonic generation}
and \emph{modulational instability} (MI), possibly resulting in
\emph{energy localization} via localized \emph{pulse formation},
is now long established in fields as diverse as Condensed Matter
Physics, Nonlinear Optics and Biophysics \cite{Davydov, Hasegawa1,
Infeld, Remoissenet}. With respect to plasma modes \cite{Krall,
Stix}, the occurrence of AM and MI has been confirmed by
experiments related to the nonlinear propagation of electrostatic
(ES, e.g. ion-acoustic) \cite{Watanabe, Bailung, Luo, Nakamura} as
well as electromagnetic (EM, e.g. whistler) waves.
Early numerical simulations of electron
cyclotron waves also predict such a behaviour \cite{Hasegawa}.

%
%
%

\begin{figure}[htb]
 \centering
\resizebox{13cm}{!}{ \includegraphics[]{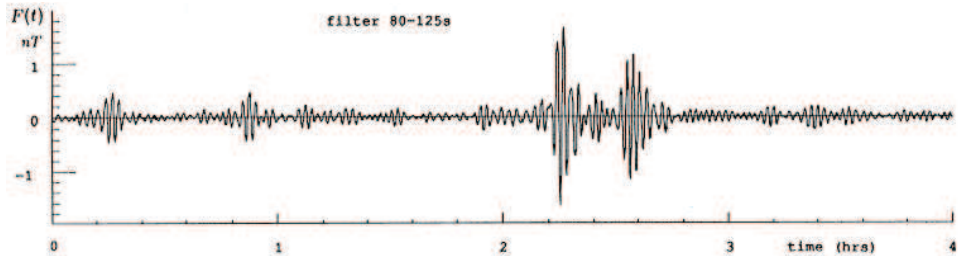} }
\caption{\small Localized envelope structures in the magnetosphere
(reprinted from \cite{Alpert}).} \label{figAlpert}
\end{figure}

\begin{figure}[htb]
 \centering
 \resizebox{13cm}{!}{
 \includegraphics[]{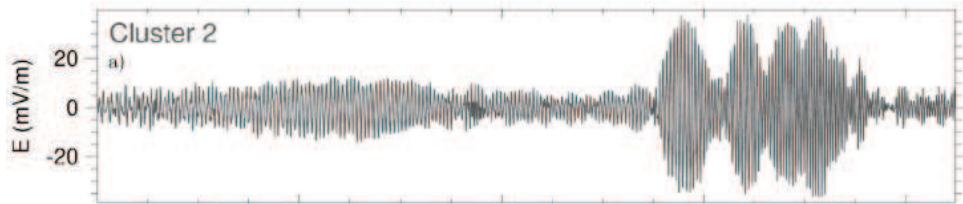}}
\caption{\small Modulated structures, related to \emph{`chorus'}
(EM) emission in the magnetosphere (CLUSTER satellite data;
reprinted from \cite{Santolik}).} \label{figSantolik}
\end{figure}
In the context of Space Physics, localized modulated wave packets
are encountered in abundance e.g. in the Earth's magnetosphere, in
fact associated with localized field and density variations which
were observed during recent satellite missions \cite{Pottelette,
Alpert, Santolik}; see e.g. Figs. \ref{figSantolik} -
\ref{figAlpert}. The occurrence of such wave forms is, for
instance, thought to be related to the broadband electrostatic
noise (BEN) encountered in the auroral region \cite{Pottelette}.
 Furthermore, recent studies have supplied evidence
for the relevance of such effects in dust-contaminated plasmas
(\emph{Dusty} or \emph{Complex} Plasmas), where a strong presence
of mesoscopic, massive, charged dust grains strongly affects the
nonlinear and dispersive characteristics plasma as a medium
\cite{Verheest, PSbook}. The modification of the plasma response
due to the presence of the dust gives rise to new ES/EM modes,
whose self-modulation was recently shown to lead to MI and soliton
formation; these include e.g. the dust-acoustic (DA) \cite{Rao,
AMS, IKPKSDAW, chin} and dust-ion acoustic (DIAW) ES modes
\cite{AMS, SSIAW, IKPKSDIAW, IKPKSEPJD}, in addition to magnetized
plasma modes, e.g. the Rao EM dust mode \cite{IKPKSRao}.
\begin{figure}[htb]
 \centering
\resizebox{9cm}{!}{ \includegraphics[]{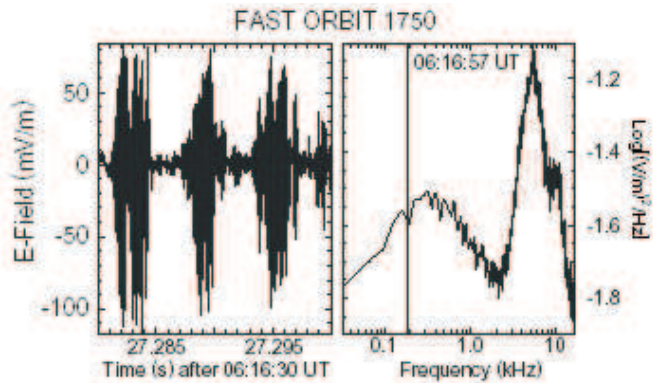} } \caption{\small
Electrostatic noise wave forms, related to modulated
electron-acoustic waves (FAST satellite data; figure reprinted
from \cite{Pottelette}). The co-existence of a high (carrier) and
a low (modulated envelope) frequencies is clearly reflected in the
Fourier spectrum, in the right.} \label{figPottelette}
\end{figure}

 The purpose of this brief study is to suggest a generic methodological
framework for the study of the nonlinear (self-) modulation of the
amplitude of electrostatic (ES) plasma modes. The general results
provided in the following are valid for various ES modes. The
generic character of the nonlinear behaviour of these modes is
emphasized, so focusing upon a  specific mode is avoided on
purpose. Where appropriate, details may be sought in the
references \cite{IKPKSDAW, IKPKSDIAW, IKPKSEPJD, IKPKSJPA,
IKPKSPRE}, where some of this material was first presented.

\section{The model: formulation and analysis}

In a general manner, several known electrostatic plasma modes
\cite{Krall, Stix} are modeled as propagating oscillations related
to one dynamical plasma constituent, say $\alpha$ (mass
$m_\alpha$, charge $q_\alpha \equiv s_\alpha Z_\alpha e$; \ $e$ is
the absolute electron charge; \ $s = s_\alpha =
q_\alpha/|q_\alpha|=\pm 1$ is the charge \emph{sign}), against a
background of one (or more) constituent(s), say $\alpha'$ (mass
$m_{\alpha'}$, charge $q_{\alpha'} \equiv s_{\alpha'} Z_{\alpha'}
e$, similarly). The background species is (are) often assumed to
obey a known distribution, e.g. to  be in a fixed (uniform) or in
a thermalized (Maxwellian) state, for simplicity, depending on the
particular aspects (e.g. frequency scales) of the physical system
considered. For instance, the \textit{ion-acoustic} (IA) mode
refers to ions ($\alpha = i$) oscillating against a Maxwellian
electron background ($\alpha' = e$) \cite{Krall, IKPKSJPA}, the
\textit{electron-acoustic} (EA) mode \cite{Krall, IKPKSPRE}
refers to electron oscillations ($\alpha = e$) against a fixed ion
background ($\alpha' = i$), and so forth \cite{Krall, Stix}. As
regards \emph{dusty plasma}  modes, the DA mode describes
oscillations of dust grains ($\alpha = d$) against a Maxwellian
electron and ion background ($\alpha' = e, \, i$) \cite{PSbook,
IKPKSDAW}, while DIA waves denote IA oscillations in the presence
of inertial dust in the background ($\alpha = i$, $\alpha' = e, \,
d$) \cite{PSbook, IKPKSDIAW, IKPKSEPJD}. Finally, this formalism
readily applies in the case when a co-existence of two different
populations of the same particle species occurs in the background,
e.g. when two different electron temperatures are present
($\alpha' = c, h$, for \textit{c}old and \textit{h}ot electrons),
affecting IA oscillations ($\alpha = i$) \cite{IKPKSJPA}; this
situation is witnessed in the upper parts of the Earth's
atmosphere.

\subsection{A generic fluid description}

A standard (single) fluid model
for  the dynamic species $\alpha$ consists of
the moment evolution equations:
\begin{eqnarray}
\frac{\partial n_\alpha}{\partial t} + \nabla \cdot (n_\alpha
\,\mathbf{u}_\alpha)&=& 0 \nonumber \\
\frac{\partial \mathbf{u}_\alpha}{\partial t} + \mathbf{u}_\alpha
\cdot \nabla \mathbf{u}_\alpha \, &=& \, -
\frac{q_\alpha}{m_\alpha}\,\nabla \,\Phi  - \, \frac{1}{m_\alpha
n_\alpha}\,\nabla p_\alpha  \nonumber
\\
\frac{\partial p_\alpha}{\partial t} + \mathbf{u}_\alpha \cdot
\nabla p_\alpha \, &=& \,  - \gamma\, p_\alpha \,\nabla \cdot
\mathbf{u}_\alpha \, ,
\end{eqnarray}
where $n_\alpha$, $\textbf{u}_\alpha$ and $p_\alpha$ respectively
denote the \textit{density}, \textit{mean} (fluid)
\textit{velocity} and \textit{pressure} of species $\alpha$.
 The electric potential $\Phi$ obeys
Poisson's eq.:
\begin{equation}
\nabla^2 \Phi \, = \, - 4 \pi \, \sum_{\alpha'' = \alpha, \{
\alpha' \}}
 q_{\alpha''} \, n_{\alpha''} \, = \,4 \pi \, e
\,(n_e  - Z_i \,n_i + ...)
\nonumber \label{Poisson}
\end{equation}
Overall charge neutrality is assumed at equilibrium, i.e.
$q_{\alpha} \, n_{\alpha, 0} = - \sum_{\alpha'}
 q_{\alpha'} \, n_{\alpha', 0}$.
The parameter $\gamma = c_P/c_V = 1 + 2/f$ denotes the specific
heat ratio (for $f$ degrees of freedom).

By choosing appropriate scales for all quantities, the above
system may be reduced to the following form:
\begin{eqnarray}
\frac{\partial n}{\partial t} + \nabla \cdot (n \,\mathbf{u})&= &
0\, , \label{reduced1}
\\
\frac{\partial \mathbf{u}}{\partial t} + \mathbf{u}  \cdot \nabla
\mathbf{u} \, &=& \, - s\,\nabla \phi \,
 - \frac{\sigma}{n}\,\nabla p \, , \label{reduced2} \\
\frac{\partial p}{\partial t} + \mathbf{u}  \cdot \nabla p \, &=
&\, - \gamma\, p \,\nabla  \cdot \mathbf{u} \,  \label{reduced3}
\end{eqnarray}
(the index $\alpha$ will be understood where omitted, viz. $s =
s_\alpha$). The re-scaled (dimensionless) dynamic variables are
now: $n = n_\alpha/n_{\alpha, 0}$, $\mathbf{u} =
\mathbf{u}_\alpha/c_*$, $p = p_\alpha/n_{\alpha, 0} k_B T_\alpha$,
and $\phi = |q_\alpha| \Phi/(k_B T_*)$, where $n_{\alpha, 0}$ is
the equilibrium density and $c_* = (k_B T_*/m_\alpha)^{1/2}$ is a
characteristic (e.g. sound) velocity. Time and space are scaled
over appropriately chosen scales $t_0$ [e.g. $\omega_{p,
\alpha}^{-1} = (4 \pi n_{\alpha, 0} q_{\alpha}^2/m_\alpha)^{-
1/2}$] and $r_0 = c_* t_0$; $T_\alpha$ is the fluid temperature,
and $T_*$ is an effective temperature (related to the background
considered), to be determined for each problem under consideration
($k_B$ is Boltzmann's constant).
The temperature ratio $T_\alpha/T_*$ is denoted by $\sigma$, in
this  \textit{warm model} \cite{warm} (the so-called \textit{cold
model} is recovered for $\sigma = 0$; see that Eq.
(\ref{reduced3}) then becomes obsolete). The Lorentz force term
was omitted, since wave propagation along the external magnetic
field is considered. The system is closed by
Poisson's equation, which may now be expressed as \footnote{A
factor $\omega_{p, \alpha}^2 t_0^2$ is omitted in the right-hand
side of Eq. (\ref{Poisson1}), since equal to 1 for $t_0 =
\omega_{p, \alpha}^{-1}$.}
\begin{equation}
\nabla^2 \phi \, =\,  - s \, \bigl[ n \, + \sum_{\alpha'}
n_{\alpha'} \, q_{\alpha'}/(n_{\alpha, 0} \, q_\alpha) \bigr] \,
\equiv -s \, (n - \hat n)\,  . \label{Poisson1}
\end{equation}
Note that the neutralizing background (reduced) density
\begin{equation}\hat n = -\sum_{\alpha'} \frac{n_{\alpha'} \,
q_{\alpha'}}{n_{\alpha, 0} \, q_\alpha} = -\, \frac{1}{s_{\alpha}
\,Z_{\alpha} \, n_{\alpha, 0}}\, \sum_{\alpha'} s_{\alpha'}
\,Z_{\alpha'} n_{\alpha'} \label{background}
\end{equation}
is \textit{a priori} \footnote{This is only not true when the
background is assumed fixed, e.g. for EA waves (i.e. $s_\alpha = -
s_{\alpha'} = -1$, $n_{\alpha'}=n_i=const.$), where $\hat n = Z_i
n_i/n_{e, 0} = const.$} a function\footnote{Note that $\hat n = 1$
for $\phi = 0$, due to the equilibrium neutrality condition. } of
the potential $\phi$; furthermore, it depends on the physical
parameters (e.g. background temperature, plasma density, defect
concentration, ...) involved in a given problem. The calculation
in the specific paradigm of IA waves is explicitly provided below,
for clarity.

\subsection{Weakly nonlinear oscillation regime}

What follows is essentially an implementation of the long known
\textit{reductive perturbation} technique \cite{redpert, Shimizu,
Kako1, Kakutani}, which was first applied in the study of electron
plasma \cite{redpert} and electron-cyclotron \cite{Hasegawa}
waves, more than three decades ago.

Equations (\ref{reduced1}) - (\ref{reduced3}) and (\ref{Poisson2})
form a system of evolution equations for the state
vector\footnote{Note that $\mathbf{S} \in \Re^{d+3}$ in a $d-$
dimensional problem ($d = 1, 2, 3$). \label{dimS}} $\mathbf{S} =
\{n, \mathbf{u}, p, \phi\}$ which accepts a harmonic
(electrostatic) wave solution in the form $\mathbf{S} =
\mathbf{S_0} \exp[i (k \mathbf{r} - \omega t)] + {\rm{c.c.}}$ Once
the amplitude of this wave becomes non-negligible, a nonlinear
harmonic generation mechanism enters into play: this is the first
signature of nonlinearity, which manifests its presence once a
slight departure from the weak-amplitude (linear) domain occurs.
In order to study the (amplitude) modulational stability profile
of these electrostatic waves, we consider small deviations
 from the equilibrium state ${\mathbf{S}^{(0)}} = (1, \mathbf{0}, 1,
0)^T$,
 viz.
 ${\mathbf{S}} = {\mathbf{S}}^{(0)} + \epsilon {\mathbf{S}}^{(1)} +
 \epsilon^2 {\mathbf{S}}^{(2)} + ...$,
 where $\epsilon \ll 1$ is a
 smallness parameter.
We have assumed that\footnote{In practice, only terms with $l \le
n$ do contribute in this summation. This simply means that up to
1st harmonics are expected for $n=1$, up to 2nd phase harmonics
for $n=2$, and so forth.} \( S_{j}^{(n)} \,= \,
\sum_{l=-\infty}^\infty \,S_{j, l}^{(n)}(X, \, T) \, e^{i l (k
\mathbf{r} - \omega t)} \) (for $j=1, 2, ..., d+3$
$^{\ref{dimS}}$; the condition $S_{j,-l}^{(n)} = {S_{j,
l}^{(n)}}^*$ holds, for reality). The wave amplitude is thus
allowed to depend on the stretched (\emph{slow}) coordinates \( X
\,= \, \epsilon (x - \lambda) \,t$ and  $T \,= \, \epsilon^2 \,
t\); \ the real variable $\lambda =
\partial \omega(k)/\partial k_x \equiv \tilde v_g$
denotes the wave's \emph{group velocity} along the modulation
direction\footnote{This is a - physically expected - constraint
which is imposed by the equations for $n=2$ and $l = 1$ (1st
harmonics at 2nd order). Alternatively, one may assume a
dependence on $X_n = \epsilon^n x$ (plus a similar expansion for
$y$, $z$ and $t$) for $n = 0, 1, 2, ...$; the condition for
annihilation of secular terms then reads: $\partial
A_1^{(1)}/\partial T_1 + (\partial \omega/\partial k_x)
\partial A_1^{(1)}/\partial X_1$, i.e. $A_1^{(1)} = A_1^{(1)}(X_1
- \tilde v_g T_1)$ (for any of the 1st harmonic amplitudes
$A_1^{(1)} \in \{ S_{1, j}^{(1)} \}$), which essentially amounts
to the same constraint.} $x$. The amplitude modulation direction
is assumed \emph{oblique} with respect to the
 (arbitrary) propagation direction\footnote{Cf. Refs. \cite{Kako,
Chhabra},
 where a similar treatment is adopted.}, expressed by the wave
vector \( \mathbf{k} = (k_x, \, k_y) = (k\, \cos\theta, \, k\,
\sin\theta) \). Accordingly, we set: $\partial/\partial t
\rightarrow \partial/\partial t - \epsilon \tilde v_g
\partial/\partial X + \epsilon^2 \partial/\partial T$, \
$\partial/\partial x \rightarrow
\partial/\partial x + \epsilon \partial/\partial X$
(while $\partial/\partial y$ remains unchanged) and \( \nabla^2
\rightarrow \nabla^2 + 2 \epsilon \, {\partial^2}/{\partial x
\partial X} + \epsilon^2 \, {\partial^2}/{\partial X^2}
\, , \) in all the above evolution equations.

By expanding near $\phi \approx 0$, Poisson's eq. may formally be
cast in the form
\begin{equation}
\nabla^2 \phi \, =\, \phi - \alpha \,\phi^2 + \alpha' \,\phi^3 - s
\,\beta\,(n - 1)\, , \label{Poisson2}
\end{equation}
where the exact form of the real coefficients $\alpha$, $\alpha'$
and $\beta$ (to be distinguished from the species indices above,
obviously) are to be determined exactly for any specific
 problem, and
 contain all the essential dependence on the plasma parameters.
 Note that the right-hand side in Eq. (\ref{Poisson2}) cancels at
 equilibrium.

\paragraph
{A case study: ion-acoustic waves}

In order to make our method and notation clear, let us explicitly
consider the simple case of \textit{ions} ($\alpha = i$ and
$q_\alpha = q_i = + Z_i e$, i.e. $s_\alpha = +1$) oscillating
against thermalized \textit{electrons} ($\alpha' = e$ and
$q_{\alpha'} = q_e = - e$, i.e. $s_{\alpha'} = - 1$; \ $n_e =
n_{e, 0} e^{e\Phi/(k_B T_e)}$). Adopting the scaling defined
above, and using the equilibrium neutrality condition $n_{e, 0} =
Z_i n_{e, 0}$, it is a trivial exercise to cast Poisson's Eq.
(\ref{Poisson}) into the (reduced) form:
\begin{equation}
\nabla^2 \phi = - (\omega_{p, i} \, r_0/c_*)^2 \,[n - e^{T_*
\phi/(Z_i T_e)}] \equiv - (n - e^{\xi \phi}) \, ,
\end{equation}
where we took: $t_0 = r_0/c_* = \omega_{p, i}^{-1}$
and $\xi \equiv T_*/(Z_i T_e)$. Now, expanding near $\phi \approx
0$, viz. $e^{\xi \phi} \approx 1 + \xi \phi + \xi^2 \phi^2/2 +
\xi^3 \phi^3/6 + ...$, we have:
\begin{equation}
\nabla^2 \phi \approx \xi \phi + \xi^2 \phi^2/2 +
\xi^3 \phi^3/6 - (n-1) \, . \end{equation} Finally, setting the
temperature scale $T_*$ equal to $T_* = Z_i T_e$, for convenience
(viz. $\xi = 1$)\footnote{Note that a different choice for $T_*$
would lead to a modified right-hand-side in Eq. (\ref{Poisson2}),
i.e. a factor $\xi \ne 1$ would precede the first
term 
(in $\phi$). This might, of course, also be a legitimate choice of
scaling; however, the following formula are not valid - and should
be appropriately modified - in this case. Obviously though, the
qualitative results of this study are not affected by the choice
of scaling. }, one recovers exactly Eq. (\ref{Poisson2}) with
$\alpha = -1/2$, $\alpha' = 1/6$ and $\beta = 1$. It may be noted
that this case has been studied, both for parallel modulation
($\theta = 0$), in Ref. \cite{Shimizu}, and for oblique
modulation, in Refs. \cite{Kako};
those results are recovered from the formulae below.

\paragraph{Amplitude evolution equations}

By substituting into Eqs. (\ref{reduced1}) - (\ref{reduced3}) and
(\ref{Poisson2}) and isolating distinct orders in $\epsilon$, one
obtains a set of reduced evolution equations in the new variables.
One is then left with the task of isolating orders in
 $\epsilon^n$ (i.e. $n = 1, 2, ...$) and successively
 solving for the harmonic amplitudes
 $S_{j, l}^{(n)}$. The calculation, particularly lengthy yet
 perfectly
 straightforward, can be found e.g. in \cite{IKPKSDIAW} for IA
 ($s=+1$) and in \cite{IKPKSPRE} for EA waves ($s=-1$).

The first harmonic amplitudes are determined (to order $\sim
\epsilon^1$)
 as
\begin{equation}
n_1^{(1)}  \,= \, s \,\frac{1 + k^2}{\beta} \,\psi \,=
\frac{k}{\omega \cos\theta}\,
 u_{1, x}^{(1)} \,= \frac{k}{\omega
\sin\theta}\,
 u_{1, y}^{(1)} = \, \frac{1}{\gamma} \, p_1^{(1)}
\label{1st-order-corrections}
\end{equation}
in terms e.g. of the potential correction $\phi_1^{(1)} \equiv
\psi$, along with the dispersion relation \( \omega^2\, = {\beta
k^2}/({k^2 + 1}) + \gamma \sigma k^2 \). Furthermore, the
amplitudes of the 2nd and 0th (constant) harmonic corrections are
obtained in $\sim \epsilon^2$; the lengthy expressions are omitted
here, for brevity \footnote{The exact expressions for the 2nd
order solution can be found e.g. in Ref. \cite{IKPKSDAW}; refer to
Eqs. (21) - (26) therein, which are exactly valid here, as they
stand.}.

\subsection
{The envelope (nonlinear Schr\"odinger) evolution equation}

The potential correction $\psi$ is found to obey
 a compatibility condition in the form of a \emph{nonlinear
Schr\"odinger--type equation} (NLSE)
\begin{equation}
i\, \frac{\partial \psi}{\partial T} + P\, \frac{\partial^2
\psi}{\partial X^2} + Q \, |\psi|^2\,\psi = 0 \, .  \label{NLSE}
\end{equation}
Both the {\em dispersion coefficient} $P$, in fact related to the
curvature of the dispersion curve as \( P \,  = \, {\partial^2
\omega}/2{\partial k_x^2} \,= \, \bigl[ \omega''(k) \,
\cos^2\theta \, + \omega'(k) \, {\sin^2\theta}/{k} \bigr]/2 \),
and the {\em nonlinearity coefficient} $Q$, which is due to
carrier wave self-interaction, are functions of $k$, $\theta$ and
$\beta$, as expected
(in addition to $\alpha, \alpha'$, for $Q$). The exact general
expressions obtained \cite{SPIG} may be \textit{tailor fit} to any
given electrostatic plasma wave problem (via the form of the
parameters $\alpha, \alpha', \beta$), in view of a numerical
investigation of the wave's amplitude dynamics (e.g. stability
profile, wave localization; see in the following).

\section{Amplitude stability profile}

It is known (see e.g. in Refs. \cite{Hasegawa1, Remoissenet,
Hasegawa-book}) that the evolution of a wave whose amplitude obeys
Eq. (\ref{NLSE}) depends on the coefficient product $P Q$, which
may be investigated in terms of the physical parameters involved.
To see this, first check that Eq. (\ref{NLSE}) supports the plane
(Stokes') wave solution $\psi = \psi_0 \exp(i Q |\psi_0|^2 T)$;
the standard linear analysis consists in perturbing the amplitude
by setting: ${\hat \psi} \, = \, {\hat \psi}_0 \, + \, \epsilon \,
{\hat \psi}_{1, 0} \cos{({\tilde k} X - {\tilde \omega} T)}$. One
thus obtains the (perturbation) dispersion relation: \( \tilde
\omega^2 \, = \, P \, \tilde k^2 \, (P \tilde k^2 \, - \, 2 {Q}
|\hat \psi_{1, 0}|^2 ) \, \). One immediately sees that if $P Q
> 0$, the amplitude $\psi$ is \emph{unstable} for $\tilde k <
\sqrt{2 {Q}/{P}}  |\psi_{1, 0}|$. If $P Q < 0$, the amplitude
$\psi$ is \emph{stable} to external perturbations.

\begin{figure}[htb]
 \centering
 \resizebox{16.5cm}{!}{
 \includegraphics{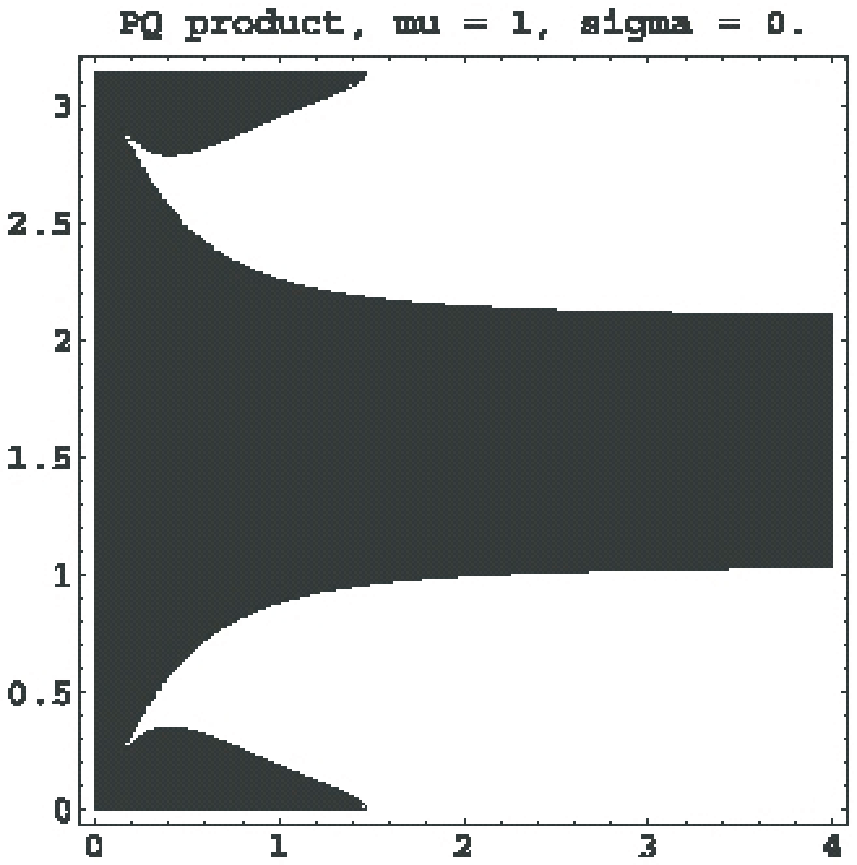}
\hskip .5 cm
\includegraphics{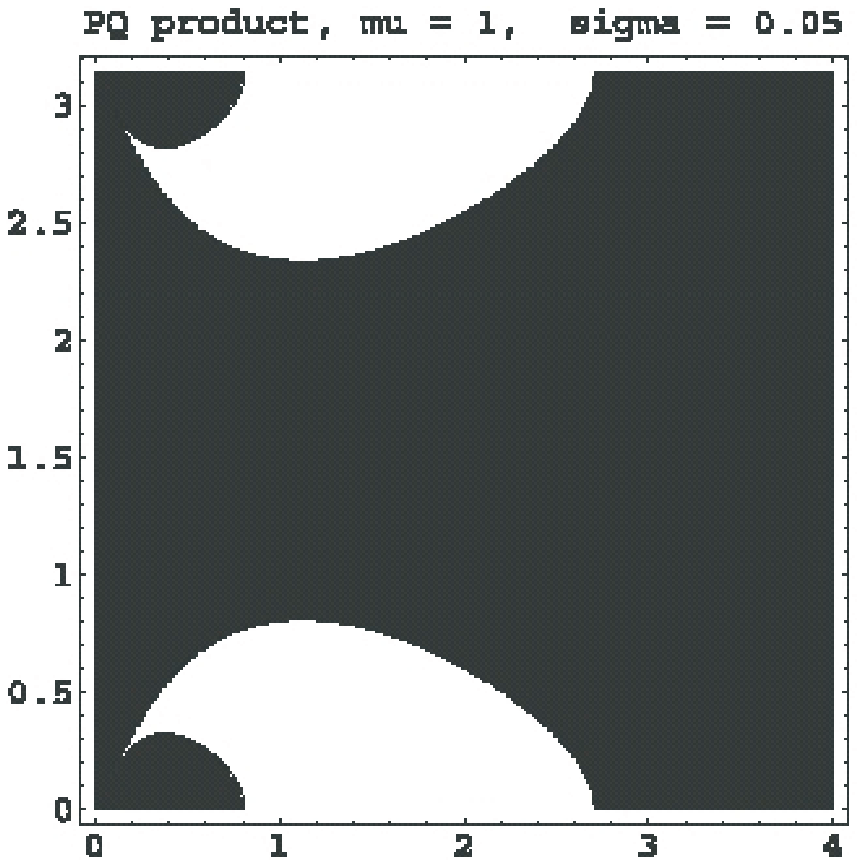}
\hskip .5 cm
 \includegraphics{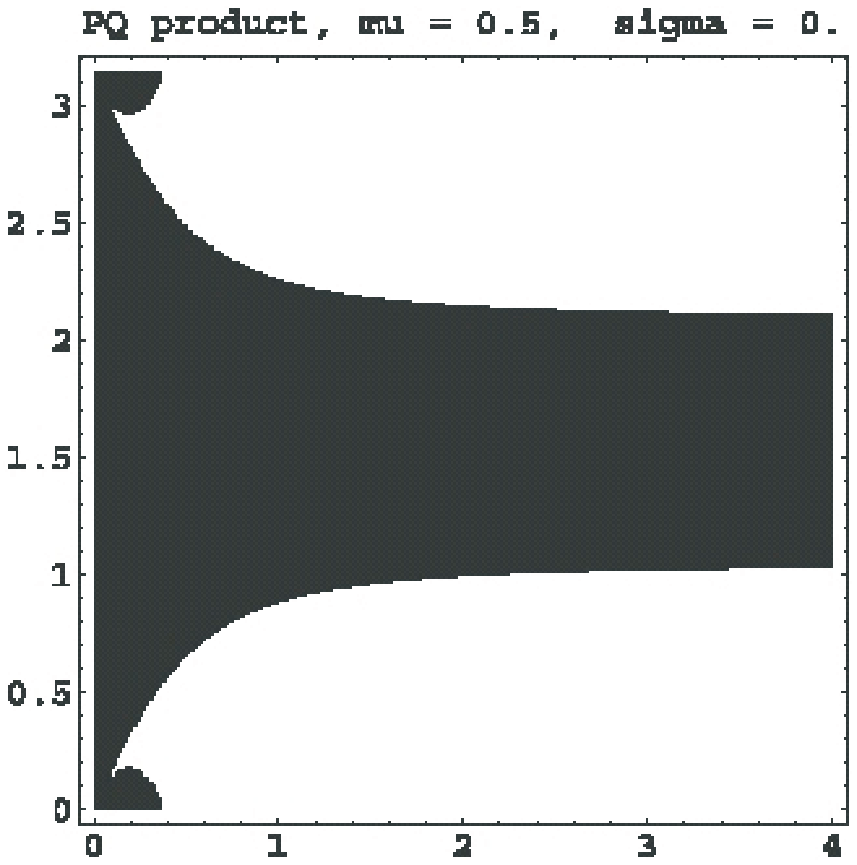}
 \hskip .5 cm
\includegraphics{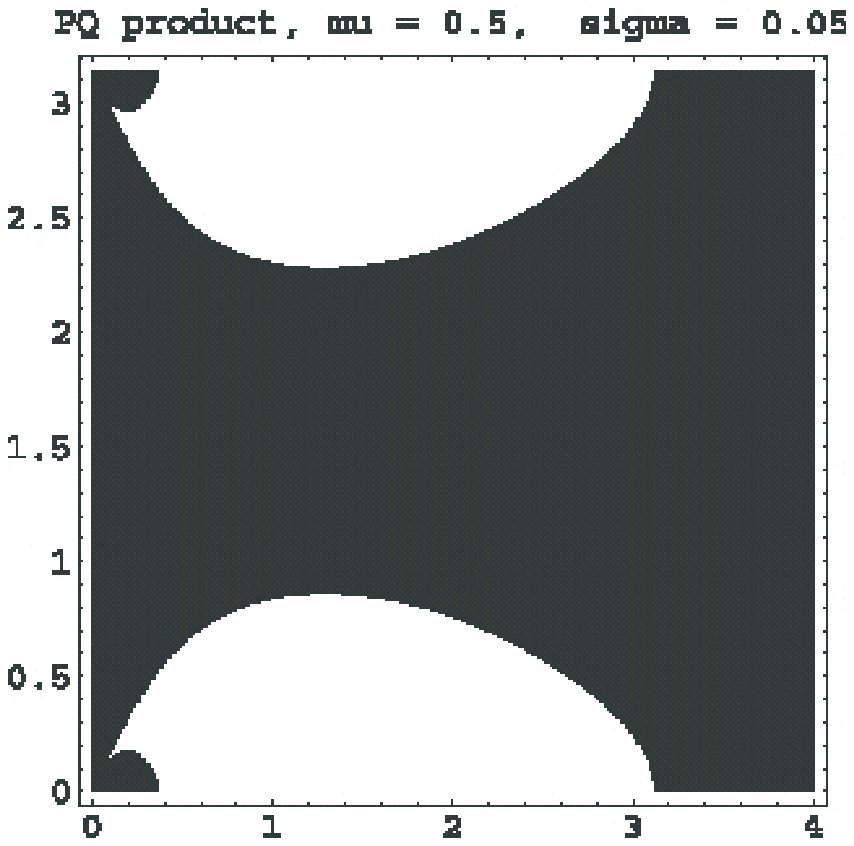}
} \caption{\small The region of positive (negative) values of the
product $P Q$ are depicted in white (black), in the
\emph{wavenumber} $k$ - \emph{modulation angle}  $\alpha$ plane.
The first two plots refer to IA waves: $\sigma = 0$ (cold model);
$\sigma = 0.05$ (warm model). Similar for the latter two, but for
DIA waves (see in the text); we have taken a \emph{negative} dust
density: $\mu = n_{e,0}/(Z_i n_{i,0}) = 0.5$
(from Ref. \cite{IKPKSEPJD}). The dust presence strongly modifies
the stability profile (rather enhancing instability here).}
\label{contour}
\end{figure}

This formalism allows for a numerical investigation of the
stability profile in terms of parameters e.g. like wavenumber $k$,
(oblique) perturbation angle $\alpha$, temperature $T_\alpha$,
background plasma parameters etc. In figure \ref{contour}, we have
depicted the region $P Q < 0$ ($P Q > 0$) in black (white) color,
for IA waves; see in Ref. \cite{IKPKSEPJD} for details.

\section{Envelope excitations}

The modulated (electrostatic potential) wave resulting from the
above analysis is of the form $\phi_1^{(1)} = \epsilon \hat \psi_0
\, \cos(\mathbf{k r} - \omega t + \Theta) + {\cal O}(\epsilon^2)$,
where the slowly varying amplitude\footnote{In fact, the potential
correction amplitude here is $\hat \psi_0 = 2 \psi_0$, from
Euler's formula: $e^{i x} + e^{-i x} = 2 \cos x$ \ ($x \in \Re$).
Note that once the potential correction $\phi_1^{(1)}$ is
determined, density, velocity and pressure corrections follow from
(\ref{1st-order-corrections}). } $\psi_0(X, T)$ and phase
correction $\Theta(X, T)$ (both real functions of $\{ X, T \}$;
see \cite{Fedele} for details) are determined by (solving) Eq.
(\ref{NLSE}) for $\psi = \psi_0 \exp(i \Theta)$. The different
types of solution thus obtained are summarized in the following.

\paragraph
{Bright-type envelope solitons}

\begin{figure}[htb]
 \centering
 \resizebox{10.cm}{!}{
 \includegraphics[]{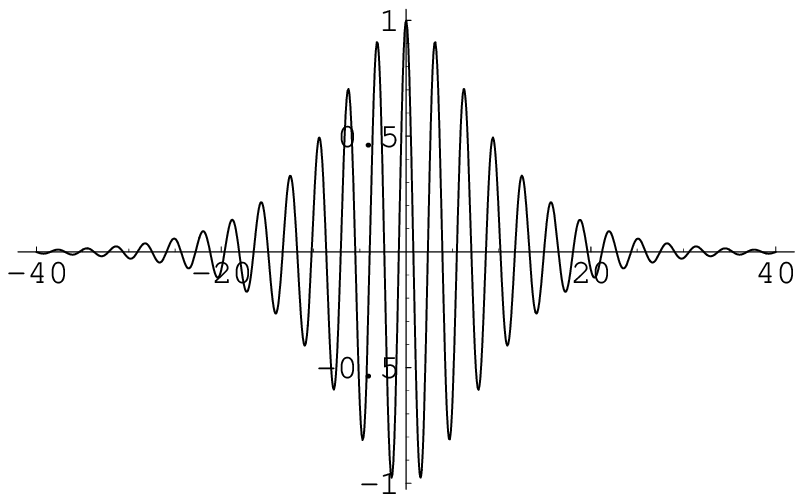}
\hskip 1.5 cm
\includegraphics{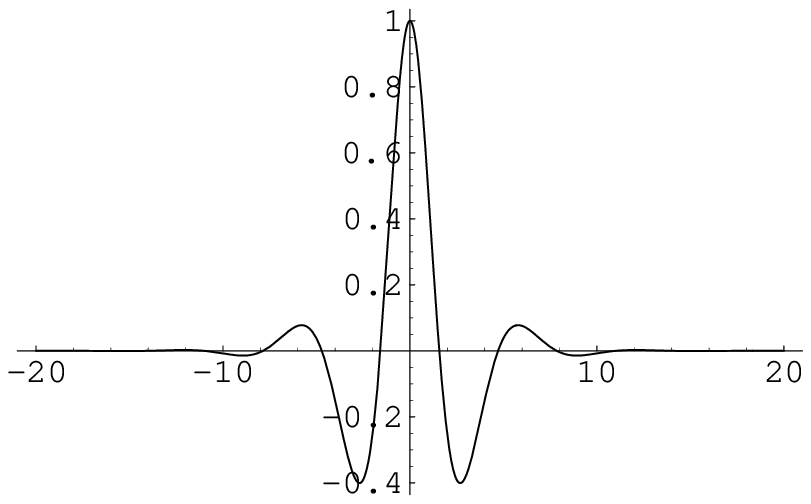}
} \caption{\small\emph{Bright}
 type modulated
wavepackets (for $P Q > 0$), for two different (arbitrary) sets of
parameter values. } \label{fig-bright}
\end{figure}

 For \emph{positive} $P Q$, the carrier wave is
modulationally \emph{unstable}; it may either \emph{collapse}, due
to external perturbations, or lead to the formation of
\emph{bright} envelope modulated wavepackets, i.e. localized
envelope \emph{pulses} confining the carrier (see Fig.
\ref{fig-bright}):
\begin{equation} \psi_0 = \biggl( \frac{2 P}{Q L^2} \biggr)^{1/2}
\, {\rm{sech}} \biggl( \frac{X - v_e \, T}{L} \biggr)\, , \quad
\Theta = \frac{1}{2 P} \biggl[ v_e  X + \biggl(\Omega -
\frac{v_e^2}{2} \biggr) T \biggr]\, \label{bright}
\end{equation}
\cite{Fedele, commentFedele}, where $v_e$ is the envelope
velocity; $L$ and $\Omega$ represent the pulse's spatial width and
oscillation frequency (at rest), respectively. We note that $L$
and $\psi_0$ satisfy $L \psi_0 = (2 P/Q)^{1/2} = {\rm{constant}}$
[in contrast with Korteweg-deVries (KdV) solitons, where $L^2
\psi_0 = {\rm{const.}}$ instead]. Also, the amplitude $\psi_0$ is
independent from the
velocity $v_e$ here.
\begin{figure}[htb]
 \centering
 \resizebox{16.8cm}{!}{
 \includegraphics{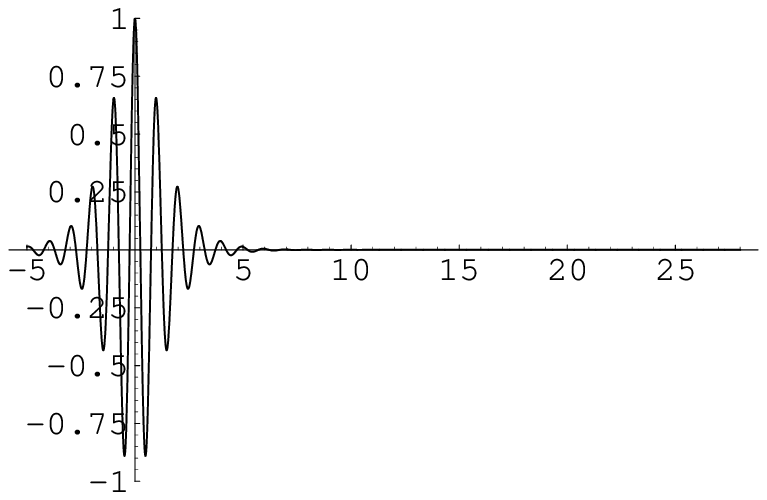}
\hskip .2 cm
\includegraphics{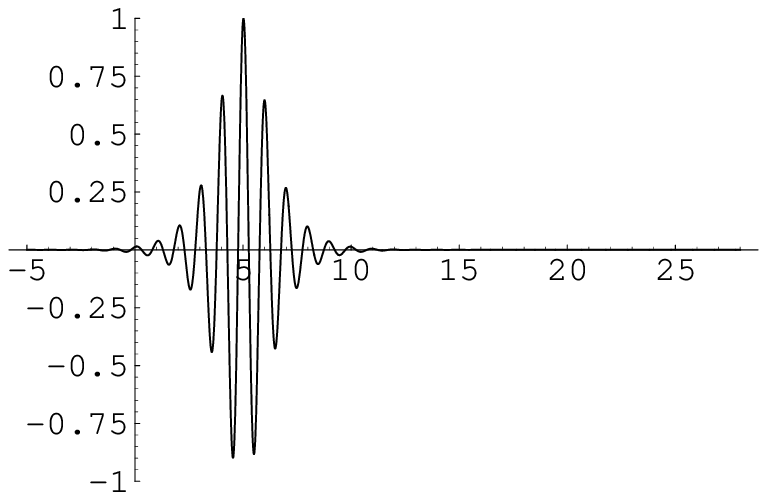}
\hskip .2 cm
 \includegraphics{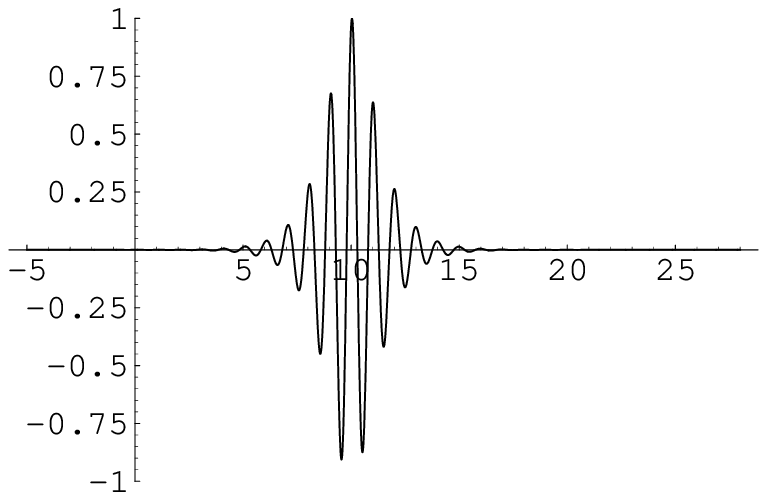}
 \hskip .2 cm
\includegraphics{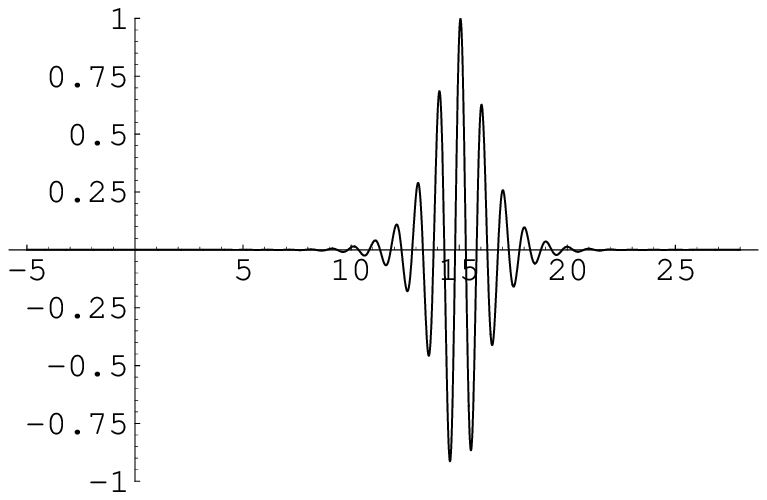}
} \caption{\small Bright envelope soliton propagation, at
different times $t_1 < \cdots < t_4$ (arbitrary parameter values):
cf. the structures encountered in satellite observations, e.g. see
Fig. \ref{figAlpert}.} \label{bright-propagation}
\end{figure}

\paragraph
{Dark-type envelope solitons}

\begin{figure}[htb]
 \centering
 \resizebox{10.5cm}{!}{
 \includegraphics[]{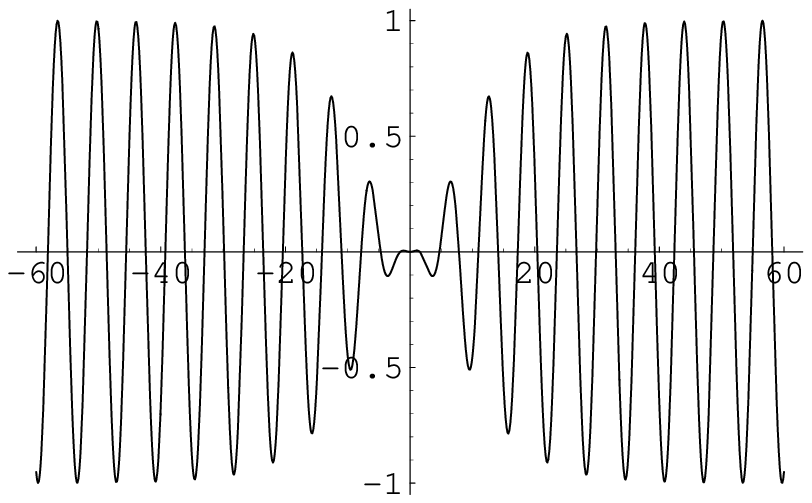}
\hskip 1.5 cm
\includegraphics{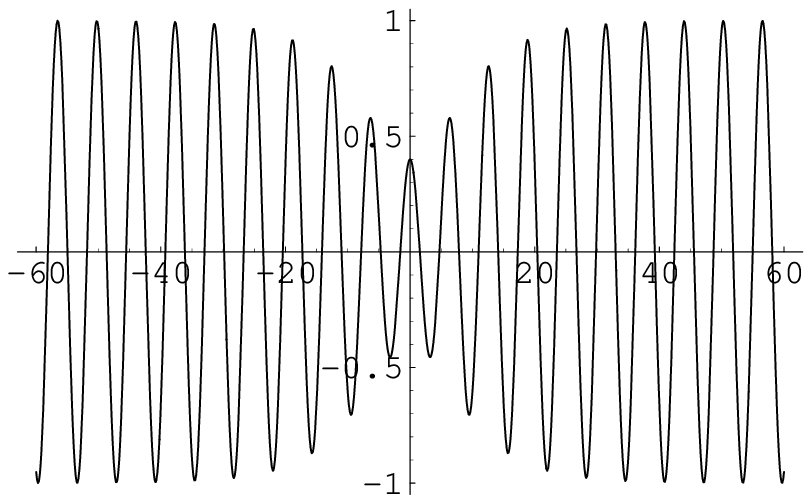}
} \caption{\emph{Dark} (left) and \emph{grey} (right) type
modulated wavepacket (for $P Q < 0$). See that the amplitude never
reaches zero in the latter case. } \label{fig-dark}
\end{figure}

For $P Q < 0$, the carrier wave is modulationally \emph{stable}
and may propagate as a
\emph{dark/grey} envelope wavepackets, i.e. a propagating
localized \emph{hole} (a \emph{void}) amidst a uniform wave energy
region. The exact expression for \emph{dark} envelopes reads
\cite{commentFedele, Fedele}:
\begin{equation}
\psi_0 = {\psi'}_0 \, \biggl| {\rm{tanh}} \biggl( \frac{X - v_e \,
T}{L'} \biggr) \biggr|\, , \qquad  \Theta =   \, \frac{1}{2 P}
\bigl[ v_e X + \bigl( 2 P Q {{\psi'}_0}^2 - \frac{v_e^2}{2} \bigr)
T \bigr] \, \label{darksoliton}
\end{equation}
(see Fig. \ref{fig-dark}a); again, $L' {\psi'}_0 = (2
|P/Q|)^{1/2}$ ($=$cst.).

\paragraph
{Grey-type envelope solitons}

The \emph{grey}-type envelope (also obtained for $P Q < 0$) reads
\cite{commentFedele, Fedele}:
\begin{eqnarray}
\psi_0 \, = \, {\psi''}_0 \, \{ 1 - d^2 \,  {\rm{sech}}^2\{[X -
v_e \, T]/L''\}\}^{1/2} \, ,  \qquad \qquad \qquad \qquad
\nonumber
\\
\Theta = \frac{1}{2 P} \, \biggl[ V_0\,X \, - \biggl(\frac{1}{2}
V_0^2 - 2 P Q {\psi''}_0^2 \biggr) \,T + \Theta_{0} \biggr]\,
- S \, \sin^{-1} \frac{d\, \tanh\bigl(\frac{X - v_e\, T}{L''}
\bigr)}{\biggr[  1 - d^2\, {\rm{sech}}^2 \biggl(\frac{X - v_e\,
T}{L''} \biggr) \biggr]^{1/2}} \, . \label{greysoliton}
\end{eqnarray}
Here $\Theta_{0}$ is a constant phase; $S$ denotes the product $S
= {\rm{sign}} (P) \, \times {\rm{sign}} (v_e - V_0)$. The pulse
width $L'' = (|P/Q|)^{1/2}/(d {\psi''}_0)$ now also depends on the
real parameter $d$, given by: \( d^2 \, = \, 1 \, + \, (v_e -
V_0)^2/({2 P Q} {{\psi''}_0^2}) \, \le \, 1 \).
$V_0 = {\rm {const.}} \in \Re$ satisfies \cite{commentFedele,
Fedele}: \( V_0 - \sqrt{2 |P Q|\, {\psi''}_0^2} \, \le \, v_e \,
\le \,V_0 + \sqrt{2 |P Q|\, {\psi''}_0^2} \). For $d = 1$ (thus
$V_0 = v_e$), one recovers the {\em dark} envelope soliton.

\section{Conclusion}

This work has been devoted to the study of the conditions for
occurrence of \emph{modulational instability}, related to the
formation of \emph{envelope localized structures}, with respect to
electrostatic waves propagating in an unmagnetized plasma. We have
shown that the envelope modulated electrostatic  wave packets
which are widely observed during satellite missions and laboratory
experiments, may be efficiently modeled by making use of a
\textit{reductive perturbation} (multiple scales) technique
\cite{redpert}. Explicit criteria are thus obtained, which
determine the wave's modulational stability profile and predict
the occurrence of localized envelope excitations of either bright
or dark/grey type. This methodology allows for an investigation of
the nonlinear modulational profile of a (any) given electrostatic
mode. Generalization in the presence of a magnetic field is on the
way and will be reported soon.


\begin{acknowledgments}
This work was supported by the {\it{SFB591
(Sonderforschungsbereich) -- Universelles Verhalten
gleichgewichtsferner Plasmen: Heizung, Transport und
Strukturbildung}} German government Programme. Support by the
European Commission (Brussels) through the Human Potential
Research and Training Network via the project entitled: ``Complex
Plasmas: The Science of Laboratory Colloidal Plasmas and
Mesospheric Charged Aerosols'' (Contract No. HPRN-CT-2000-00140)
is also acknowledged.
\end{acknowledgments}




\begin{thebibliography}{10}
\bibitem{commentFedele} These expressions are readily obtained
from Ref. \cite{Fedele}, by shifting the variables therein to our
notation as: $x \rightarrow \zeta$, \, $s \rightarrow \tau$, \,
$\rho_m \rightarrow \rho_0$, \ $\alpha \rightarrow 2 P$, \ $q_0
\rightarrow - 2 P Q$, \ $\Delta \rightarrow L$, \ $E \rightarrow
\Omega$, \ $V_0 \rightarrow u$.
\bibitem{comment2} All the expressions from (9) and on in in
Ref. \cite{IKPKSDAW} are exactly valid here; this is due to the
generic form of the reduced Eqs. (\ref{reduced1}) -
(\ref{reduced3}) and (\ref{Poisson2}) of the model.
\end{thebibliography}
\end{document}